\documentclass[12pt,preprint]{aastex}
\usepackage{emulateapj5}
\usepackage{apjfonts}

\submitted{Accepted for Publication in ApJ Letters}

\shorttitle{Jet Collimation in Action}
\shortauthors{Homan et al.}

\begin{document}

\title{Jet Collimation in Action: Re-alignment on Kiloparsec Scales in
3C\,279}
\author{D. C. Homan\altaffilmark{1}, M. L. Lister\altaffilmark{1}, K. I. Kellermann\altaffilmark{1},
M. H. Cohen\altaffilmark{2}, E. Ros\altaffilmark{3},
J. A. Zensus\altaffilmark{3}, M. Kadler\altaffilmark{3}, R. C. Vermeulen\altaffilmark{4}}

\altaffiltext{1}{National Radio Astronomy Observatory,
Charlottesville, VA 22903; dhoman@nrao.edu, mlister@nrao.edu; kkellerm@nrao.edu}

\altaffiltext{2}{Department of Astronomy, MS 105-24, California
Institute of Technology, Pasadena, CA 91125; mhc@astro.caltech.edu}

\altaffiltext{3}{Max-Planck-Institut f\"ur Radioastronomie, Auf dem H\"ugel 69,
D-53121 Bonn, Germany; ros@mpifr-bonn.mpg.de,
azensus@mpifr-bonn.mpg.de, mkadler@mpifr-bonn.mpg.de}

\altaffiltext{4}{Netherlands Foundation for Research in Astronomy, 
Postbus 2, NL-7990 AA Dwingeloo, Netherlands; rvermeulen@astron.nl}

\vspace{0.15in}

\begin{abstract}
We report a change in the trajectory of a well studied jet component 
of the quasar 3C\,279.  The component changes in apparent projected speed 
and direction, and we find it to be moving with a Lorentz factor 
$\gamma \gtrsim 15$ at an initial angle of $\lesssim 1^\circ$ to 
the line of sight. The new trajectory of the component has
approximately the same speed and direction as an earlier 
superluminal feature, originally seen in the early 1970s.  The new
direction for the component is also much better aligned with larger 
scale VLBA and VLA structure out to $0.1\arcsec$.  
We suggest that the trajectory change is a collimation event 
occurring at $\gtrsim 1$ kiloparsec (deprojected) along the jet.  
While the change in trajectory on the sky appears to be $26^\circ$ the 
intrinsic change is $\lesssim 1^\circ$.  We estimate the Doppler factor 
prior to the change in direction to be $\delta \gtrsim 28$ and after 
the change to be $\delta \gtrsim 23$.  Comparison to independent
constraints on the Doppler factor suggest that the energy in the
radiating particles cannot greatly exceed the energy in the magnetic
field unless the volume filling factor is very much less than one.
\end{abstract}


\vspace{0.05in}

\section{Introduction}
\label{s:intro}

Faster-than-light or ``superluminal'' motion was originally predicted
by \citet{R66} and first observed by \citet{W71} and \citet{C71} 
in the powerful radio quasar 3C\,279 (1253$-$055; $z=0.536$). The illusion 
of superluminal speed results from highly relativistic motion directed 
toward the observer, with the moving material chasing the radiation it 
emits, creating a compression in the time sequence of events as seen by 
the observer.  In radio quasars, like 3C\,279, a pair of highly 
collimated, relativistic plasma jets stream outward from the nucleus 
or core, and if one of these jets points within a small angle, $\theta$, 
of our line of sight, then pattern motion with some intrinsic speed, 
$\beta = v/c < 1$, will appear to move across the sky with an apparent 
speed given by:

\begin{equation}
\beta_{app} = \frac{\beta\sin\theta}{1-\beta\cos\theta}
\end{equation} 
which can greatly exceed unity.

The radio jet in 3C\,279 is a particularly interesting example with
components moving along multiple position angles and apparent speeds.  The
motion of the original superluminal feature, as summarized by \citet{C79},
who analyzed the six available epochs spanning the period 
1970--1972, was along a position angle of $-142^\circ\pm2^\circ$ 
at $0.5\pm0.1$ mas/yr.  For a modern cosmology of 
$H_0=70$\,km\,s$^{-1}$\,Mpc$^{-1}$,
$\Omega_M = 0.3$, and $\Omega_\Lambda = 0.7$, this corresponds to 
$\beta_{app} = 16\pm3$.  The VLA observations of \citet{dPP83} showed
that knot C at $0.1\arcsec$ had a position angle of $-145^\circ$ in
a nearly direct line with the superluminal motion detected on length
scales fifty times smaller.
The observational picture became less clear during the 1980s
when \citet{U89} tracked another superluminal component
from 3C\,279 which they called ``C3''.  Component C3 was found to move 
along a different position angle of $-134^\circ$ with a much smaller 
speed of only $0.12\pm0.02$ mas/yr corresponding to $\beta_{app} = 4\pm1$.  

At the very end of 1984, \citet{U89} note the emergence of a new
component they call C4.  Component C4 is an extremely strong and
compact feature which has been tracked up to the present by
a number of programs.  The most detailed and complete
results to date were presented by \citet{W01}.  They followed
C4 from 1991 to the end of 1997 with 22 GHz observations and from 1995
to 1997 at 43 GHz.  They found the component to be moving
along a position angle of $-114^\circ$ with a speed of $0.26\pm0.01$ 
mas/yr which corresponds to $8.2\pm0.3$ times the speed of light in our
choice of cosmology.  They extrapolate an origin epoch for the
component of $1984.7\pm0.3$, consistent with the first observation
of C4 by \citet{U89}.  \citet{W01} discuss a dip in speed of C4 
around 1994 which coincides with small changes
(a few degrees) of the trajectory on the sky.  Whether this small
change in the motion is related to the large changes reported
here is unclear; however, the component had resumed its previous 
motion and trajectory by the end of 1995. 

Here we present results from the Very Long Baseline Array\footnote{
The VLBA is operated by the National Radio Astronomy Observatory which
is a facility of the National Science Foundation operated under 
cooperative agreement by Associated Universities, Inc.} (VLBA) 
2cm Survey \citep[\url{http://www.cv.nrao.edu/2cmsurvey}]{K98} showing 
that after 1998, the trajectory of C4 changes in both apparent speed
and direction.  The new trajectory for C4 is on an essentially 
parallel track to the 1970s superluminal component with essentially 
the same speed as originally observed for that component.  

\begin{center}
\figurenum{1}
\plotone{f1.eps}
\figcaption[f1.eps]{\label{f:pos}
A $\lambda$2 cm, VLBA image of 3C\,279 in March 1997.
Contours begin at 10 mJy/beam and increase in powers of two.  The
cross in the NW corner of the map depicts the FWHM dimensions of the 
restoring beam. 
Overlayed on this image in red are component positions for C4. 
Solid triangles are 11 GHz data from \citet{C93}, open squares are 
22/43 GHz data from \citet{W01} (pre 1995.5), and solid circles are 
measured from 2 cm VLBA data taken from the period 1995.5 to 2002.9.  
Blue open stars and green open crosses indicate the approximate
locations measured for the 1970s superluminal component and
component C3 respectively (distances and position angle taken from
\citet{C79} and \citet{U89}).  Red, green, and blue dashed lines mark 
the $-114^\circ$, $-134^\circ$, and $-142^\circ$ position angles respectively.
}
\vspace{0.1in}
\end{center}

\section{Observations and Results}
\label{s:obs}

Our observations of 3C\,279 from the VLBA 2cm Survey consist of 12
epochs spanning the interval from 1995.5 to 2002.9 
(for an animation
see our project website referenced above).  We have also
incorporated 2cm VLBA results from the Brandeis University
parsec-scale jet monitoring project \citep[Ojha et al. in
prep.]{H01,H02} which adds six closely
spaced epochs during 1996 and a seventh epoch at the end of 1997.
We measured component positions both in the image plane, by fitting
point sources at the peak locations of the core and C4, and in the 
$(u,v)$-plane by fitting 2 or 3 point components to the core region and
a single elliptical Gaussian to C4.  We have averaged the values 
obtained by these two approaches, and have plotted the mean position in 
figures \ref{f:pos} and \ref{f:motion}.  The core is assumed to be
stationary, and the positions of C4 are taken relative to the core. 

Figure \ref{f:pos} shows an image of 3C\,279 from our March 1997 
observations.  Superimposed on this image is the position of 
component C4 from each of the 19 epochs we compiled.  We have also 
plotted
22/43 GHz data from \citet{W01} up to 1995.5 and 11 GHz data from
\citet{C93} to illustrate the trajectory of C4 from 1988 to the 
present.  Approximate positions for the 1970s superluminal component 
and the component C3 are included on the figure for comparison.  

Proper motions were computed from the 2cm $(x,y)$ compo-
\begin{center}
\figurenum{2}
\plotone{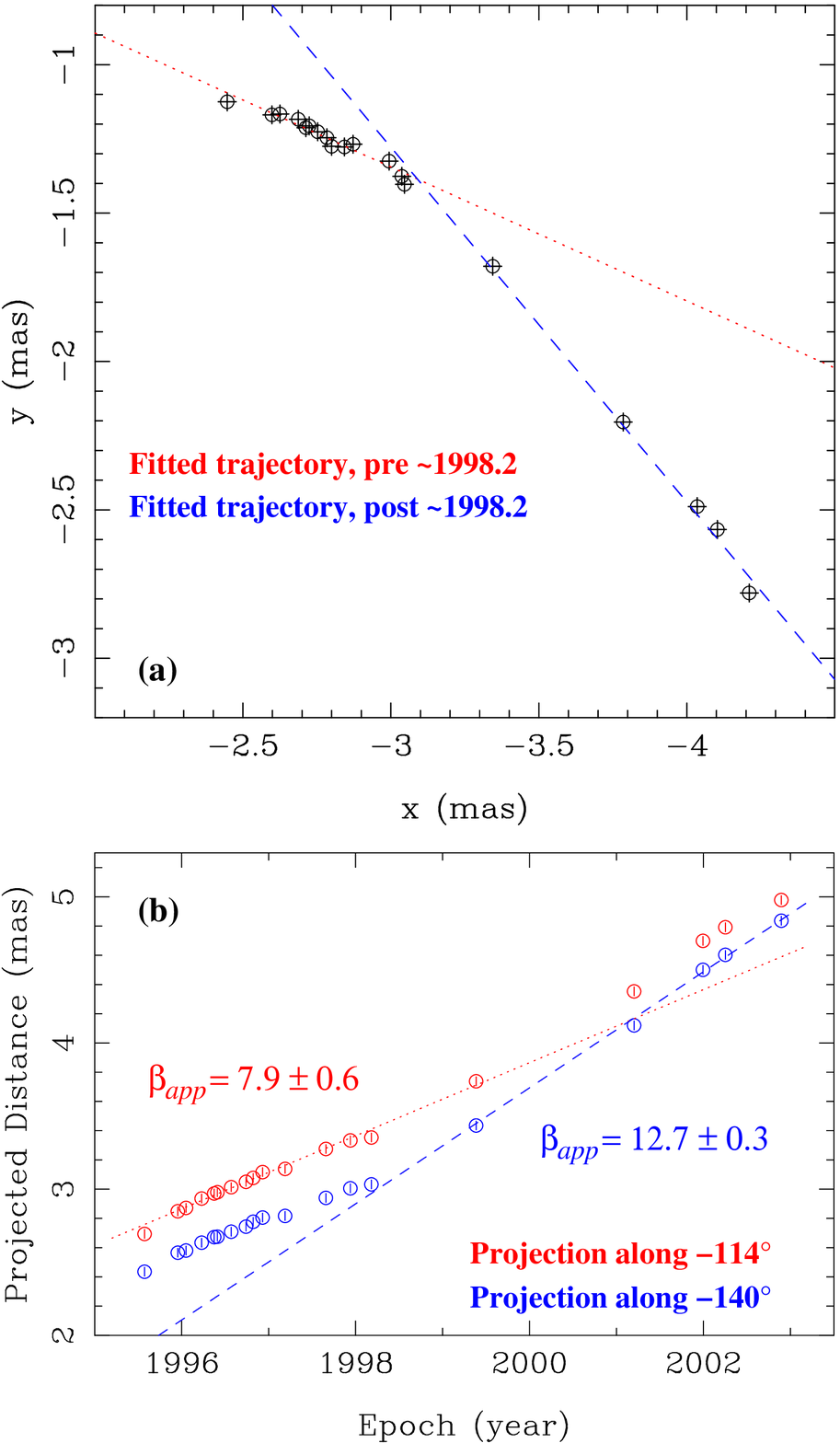}
\figcaption[f2.eps]{\label{f:motion}
Component locations and fitted proper motions of C4 before and
after the bend.  Panel (a) plots the component locations and fitted 
motions as seen on the sky.  Panel (b) plots the component distances
and fitted motions versus time.  The distances in panel (b) are 
projected distances along the position angles corresponding to the 
directions of motion before and after the bend.
}
\vspace{0.1in}
\end{center}
nent 
positions in both the image and $(u,v)$ planes independently.  
The motions were computed over two different intervals, one prior to the 
change in trajectory and one after.  We were unable 
to decide 
whether epoch 1998.2 should be classified with the 
motion prior to 
the change or after the change, but we found that it made little 
difference which choice we made.  The average motions we found
across all of these approaches were the following: 
{\em prior to $\sim$1998.2}: $\mu=0.25\pm0.02$ mas/yr ($7.9\pm0.6$
$c$) along a 
position angle of $-114^\circ\pm1^\circ$, and {\em after $\sim$1998.2}:
$\mu=0.40\pm0.01$ mas/yr ($12.7\pm0.3$ $c$) along PA$=-140^\circ\pm1^\circ$.
These proper motions are plotted against the data in figure 
\ref{f:motion}.  We estimate our typical positional uncertainty 
for the centroid of C4 to be $0.03$ mas in both dimensions.

Figure \ref{f:flux} depicts the change in component flux density
(panel (a)) and size along the $-114^\circ$ position angle (panel (b))
as a function of epoch.  Prior to the sharp change in motion sometime
in 1998, the component increases in flux density by $50$\% 
\begin{center}
\figurenum{3}
\plotone{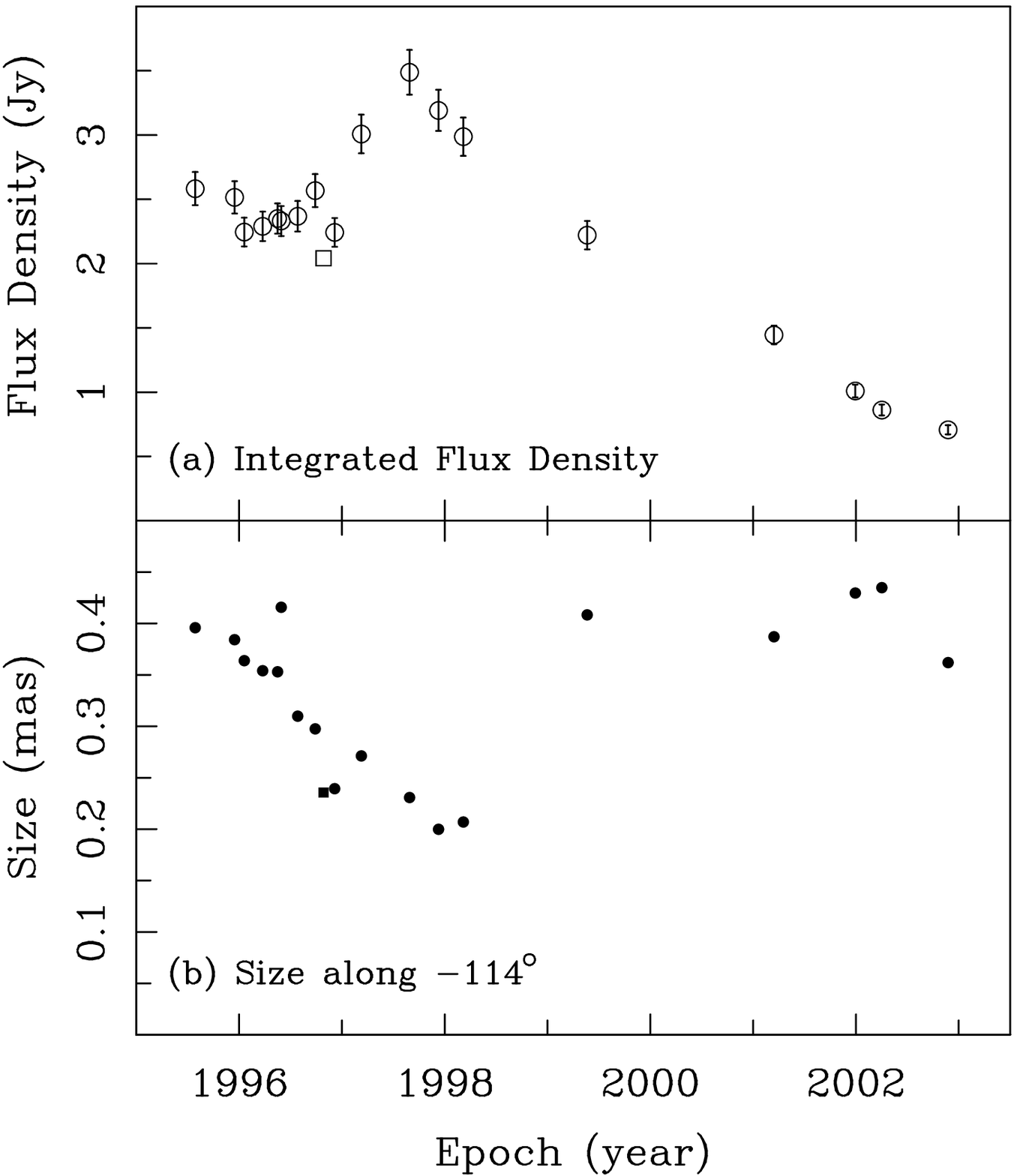}
\figcaption[f3.eps]{\label{f:flux}
Plots of component C4's flux density (panel (a)) and FWHM angular 
size projected along the $-114^\circ$ position angle (panel (b)).  
These properties are taken from our Gaussian fit to the (u,v)-plane
data.  The square symbols represent a poorly calibrated epoch 
with flux density $\sim 10$\% low.  Error bars on the flux density 
points are estimated at 5\% of the integrated flux density in each 
epoch (e.g. \citet{H02}).  
}
\vspace{0.1in}
\end{center}
and 
decreases in size along the direction of motion by a factor of two.  
These events seem to come before the change in motion, suggesting the 
component is already undergoing changes, 
likely caused by interactions 
with its environment, while it is 
still propagating along the original 
direction of motion.  \citet{ZT01} report a change in the polarization 
position angle of C4 which occurred sometime between 1998.6
\citep{T00} and 2000.1.  The polarization angle change reflects a 
change in magnetic field order, as Faraday rotation is small in this
feature \citep{ZT01}, and provides further evidence that the change 
in trajectory is tied to internal changes in component structure.

\section{Analysis}

Here we make a simple ``billiard-ball'' analysis of the component's 
trajectory.  We note that more complicated models are possible,
perhaps involving very different pattern and flow speeds, and such
models may allow a wider range of allowed $\gamma$ and $\theta$.

The apparent speed of this component has increased by more
than $50$\% and the trajectory on the sky has changed by $26^\circ$.
According to equation 1, the apparent speed can increase either due
to an increase in the intrinsic speed, $\beta$, or by a change
in angle to the line of sight. We observe a large change in 
projected angle on the sky, therefore it is reasonable to assume that 
the angle to the line of sight has also changed.  We cannot rule out 
an increase in beta; however, the simplest assumption consistent with
our observations is that the speed of the component is constant along 
a bent path, and we make that assumption here.

Figure \ref{f:doppler} plots the Doppler factor, $\delta =
1/\gamma(1-\beta\cos\theta)$ where 
\begin{center}
\figurenum{4}
\plotone{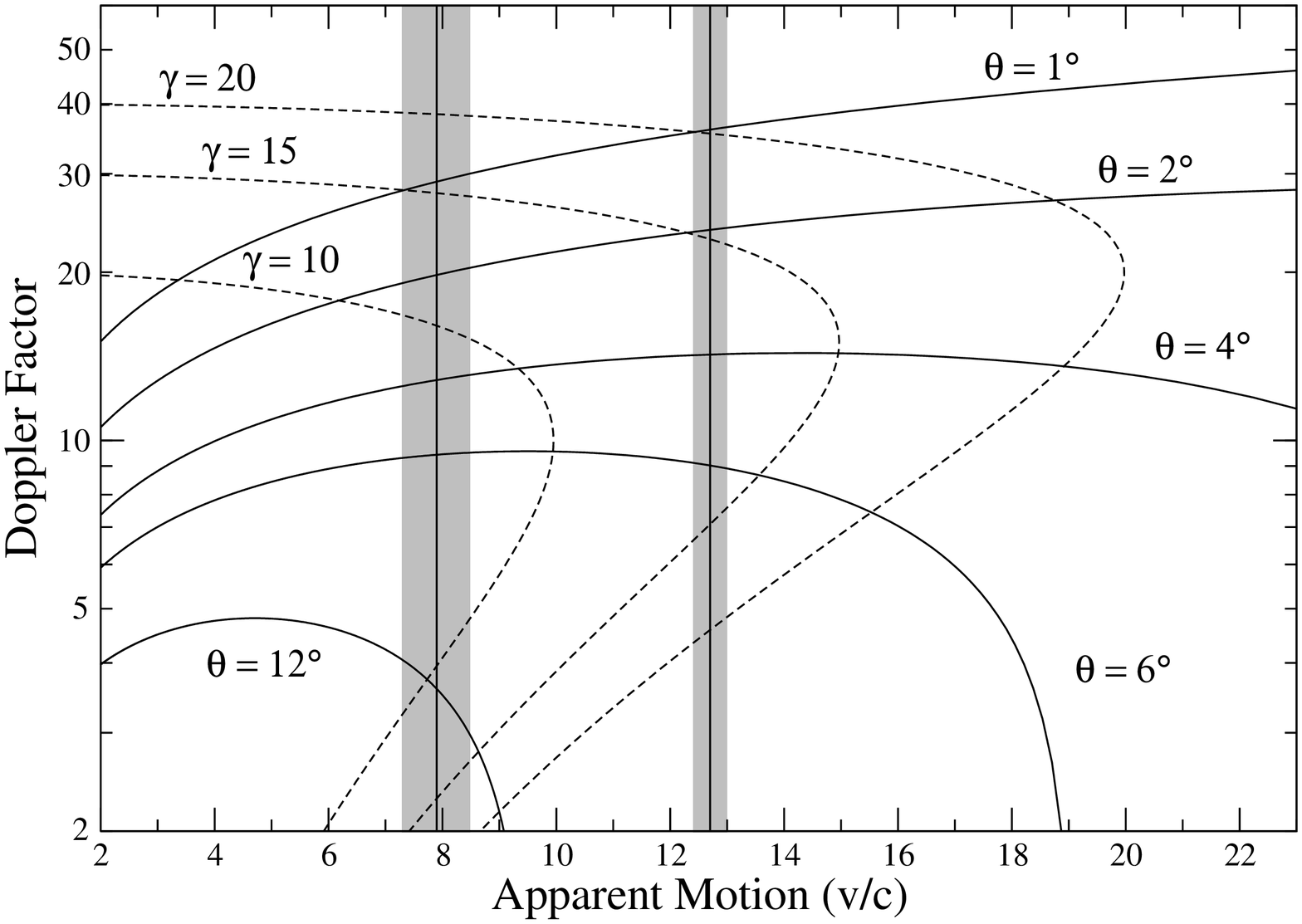}
\figcaption[f4.eps]{\label{f:doppler}
Plot of Doppler factor versus apparent speed.  Dashed contours indicate 
lines of constant bulk Lorentz factor, and solid contours indicate
lines of constant angle to the line of sight.  Vertical lines with
gray envelopes indicate the component speed before ($7.9\pm0.6$ c) and
after ($12.7\pm0.3$ c) the bend. 
}
\vspace{0.1in}
\end{center}
$\gamma = 1/\sqrt{1-\beta^2}$,
versus the apparent motion, $\beta_{app}$.  Lines of constant speed
(Lorentz factor, $\gamma$) and lines of constant angle are shown
on the figure.  $\gamma = 10$ is clearly too small and cannot
accommodate speeds greater than $10c$.  The lines for $\gamma=15$
or $\gamma=20$ are more reasonable, and the change in apparent speed 
from $8c$ to $13c$ can be accommodated by either a small increase,
$\sim 1^\circ$, or a larger decrease, $\sim 6^\circ$, in the angle 
of the motion with respect to the line of sight.

To decide between the two possibilities of increasing or decreasing
$\theta$, we must look at the corresponding change in Doppler factor
which will scale the flux density of the jet component like $S \propto
\delta^{3.5}$, appropriate for a simple discrete component with a spectral
index of $0.5$.   For a decreasing $\theta$ (i.e. a
trajectory bent towards the line of sight), the Doppler factor more
than doubles between $8c$ and $13c$ along the $\gamma=15$ curve.
This would predict an increase in flux density of the component of more
than an order of magnitude; however, from figure \ref{f:flux}a, it is clear
that the flux density of the component changes by no more than a factor 
of two, and after 1998 the flux density only decreases.  Note that 
the observed compression in component size prior to the bend would
be likely only to further increase the flux density, and therefore with some
confidence we reject a decreasing angle to the line of sight.  For increasing
$\theta$ (i.e. a trajectory bent away from the line of sight) the 
situation is much better as the Doppler factor decreases by
approximately $20$\% along the $\gamma=15$ curve.  This corresponds to
a decrease in component flux density of a factor of two, somewhat more
than we actually observe but not unreasonably so.  

Considering the complex nature of the light curve which is influenced
not only by Doppler beaming but also by the observed compression of
the component prior to the bend and expansion after the bend, we
don't place strict limits here on how the Doppler factor has 
decreased after the bend.  While the $\gamma=15$ curve seems to
provide a reasonable lower bound on C4's Lorentz factor, we
note that the Lorentz factor could be much larger and still be 
consistent with our observations.

From this analysis assuming fixed component speed, we estimate
$\gamma \gtrsim 15$ with an initial angle to the line of sight of 
$\theta \lesssim 1^\circ$ increasing at the bend to become 
$\theta \lesssim 2^\circ$. The bend in the plane of the sky appears to
be $26^\circ$; however, deprojected, this bend is only 
$\sim 26^\circ \times \sin\theta \sim 0.5^\circ-1^\circ$, consistent 
with the required bend along the line of sight.  

\section{Discussion}

\subsection{The Nature of Component C4}

In the above analysis, we have taken a very simple ``billiard-ball''
model of the motion of component C4, where C4 undergoes a sudden
change in trajectory sometime during 1998.   With a major axis size 
(FWHM) of $\lesssim 10$ ly (0.5 mas) and a Doppler factor $> 20$, it 
is not unreasonable that C4 could have changed its trajectory in less 
than a year in our frame, and while this simple model fits 
our data quite well, a more detailed analysis of the trajectory
is desirable.  Unfortunately we are limited in this by a lack
of good sampling in the period 1998--2000.  An additional 
problem for more detailed analysis of the trajectory 
is that C4 is an extended region, and we are only tracking its 
centroid.  In reality, C4 may be a complex shocked region with a 
strong leading edge and trailing shocks, and small changes in its 
centroid may not reflect true kinematical changes. In this light, 
we believe our simple approach is a conservative one that leads 
to a consistent picture of the jet kinematics.

\subsection{Doppler Factor and Equipartition}

From the angles and Lorentz factors estimated above, we can derive
approximate limits on the Doppler beaming factor of this jet feature 
before and after the bend.  We find $\delta \gtrsim 28$ prior to 
the change in trajectory and $\delta \gtrsim 23$ after the change.  
These values are consistent with \citet{P03}, who use an
inverse-Compton calculation to estimate an average lower limit for 
the Doppler factor of C4, $\delta > 21$, prior to the change in 
direction.

It is interesting to compare our values to the completely
independent estimate of the Doppler factor made by \citet{HW00}
assuming equipartition between the radiating particles and magnetic
field.  They observed a spectral turnover at $6$ GHz for C4
in epoch 1997.94, just before the change in trajectory 
described here.  They used the spectrum and resolved
angular size of C4 to calculate the Doppler factor:  
$\delta = 18^{+6}_{-3}(\eta/\Upsilon)^{1/7}$ where $\eta$ is the 
ratio of energy in the magnetic field to energy in the radiating 
particles: $\eta = U_B/U_p$, and where $\Upsilon \leq 1$ is the
fraction of the volume filled with radiating plasma\footnote{We
adjusted Homan \& Wardle's result for our choice of cosmology and 
inserted the dependence on $\Upsilon$.}.  Agreement between these two 
independent determinations of the Doppler factor of component C4 
requires $\eta/\Upsilon \gtrsim 1$, implying that the energy in the 
radiating particles cannot greatly exceed the energy in the magnetic 
field unless the volume filling factor is very much less than unity.

\vspace{0.5in}

\subsection{What Caused the Bend?}

We argue above that C4 was interacting with its environment prior to 
the change in trajectory, resulting in decreased angular size along
the $-114^\circ$ position angle and increased flux density.  It seems
likely, therefore, that C4 has been deflected onto its new trajectory.
A random collision would be unlikely to deflect C4
in just the right fashion to have essentially the same speed
{\em and} direction on the sky as the superluminal component 
detected in 1971.  It is important to stress that while the trajectory 
of C4 is {\em parallel} to that of the older component, C4 is
not following the actual path of that component (see figure 1).
The new trajectory for C4 not only has a very similar speed and
direction as the 1970s superluminal feature, but it is also in the 
direction of the larger scale structure as seen in the 1.6 GHz VLBA
maps made by \citet{P00}.

We suggest that the change in trajectory is a collimation event 
resulting from the interaction of C4 with the boundary between the 
jet outflow and the interstellar medium.  The exact nature of this
boundary and interaction is unclear; however, it is interesting to
note that (a) the apparent collimation event occurred at a deprojected 
distance from the nucleus of $\gtrsim 1$ kpc, and (b) the jet is still 
highly relativistic, with $\gamma \gtrsim 15$, on these scales.

\citet{AC98} discuss a precessing jet model for 3C\,279 to 
explain the different apparent speeds and position angles of 
the VLBI components appearing prior to 1990.  \citet{W01} confirm
the presence of different apparent speeds and position angles in
the jet, even in newer components originating after C4, although their 
measurements do not agree with the extrapolated predictions of the 
Abraham and Carrara model.  The concordance of the new speed for C4 
with that of the 1970s component adds weight to the idea that jet 
components may all have intrinsically similar Lorentz factors 
but slightly different angles of ejection.  If some sort of precession
model is used to interpret the data, Wehrle et al. comment that it is 
difficult to understand how the range of ejection angles from $-114^\circ$
to $-142^\circ$ can be consistent with the narrow range of angles
observed in the larger scale VLBA and VLA structure.  Perhaps we have 
seen the answer here in the form of jet collimation that occurs at
$\sim 1$ kpc or more.

\acknowledgments

We thank R. Ojha, D. Roberts, and J. Wardle for sharing their 
monitoring data.  J.~A.~Z. was supported for this research through 
a Max-Planck Research Award. This research has made use of data from 
the University of Michigan Radio Astronomy Observatory, which is 
supported by the National Science Foundation and by funds from the 
University of Michigan.



\end{document}